\documentclass{kapproc} 
\usepackage{epsfig}
\newcommand{\be}{\begin{equation}}
\newcommand{\ee}{\end{equation}}

\begin{document}

\articletitle{Positional correlation between low-latitude
gamma-ray sources and SNRs}

\author{Diego F. Torres, Jorge A. Combi, Gustavo E. Romero
and P. Benaglia} \affil{Instituto Argentino de
Radioastronom\'{\i}a\\ C.C. 5, (1894) Villa Elisa, Buenos Aires\\
Argentina} \email{dtorres@venus.fisica.unlp.edu.ar}



\begin{abstract}
We present the results of a spatial correlation analysis between
unidentified gamma-ray sources in the 3EG catalog and candidates
to supernova remnants. This work extends a previous study made by
Romero, Benaglia \& Torres (1999). We also consider the gamma-ray
emission variability and the spectral index for the sources with
positional coincidence.
\end{abstract}

\section{SNRs as gamma-ray emitters}

Supernova remnants (SNRs) are usually considered as the main
sources of cosmic rays with energies below $\sim 10^{15}$ eV. Both
electrons and protons are believed to be accelerated by Fermi
mechanism in the expanding shock front of these objects. The
electrons produce synchrotron emission detected at radio
wavelengths whereas the interactions between relativistic protons
and ambient nuclei can produce neutral pions, which quickly decay
yielding $\gamma$-ray emission at energies $E \geq 100$MeV, in the
EGRET range. The expected $\gamma$-ray flux at Earth was given by
Drury et al. (1994), \be F{(\geq 100 {\rm MeV})} \sim 4.4 \times
10^{-7}  \theta \frac {E_{SN}}{10^{51} {\rm erg}}
\left(\frac{d}{{\rm kpc}} \right) ^{-2} \left(\frac{n}{{\rm
cm}^{-3}} \right) {\rm cm}^{-2} {\rm s}^{-1},\ee where $E_{SN}$ is
the energy of the SN in ergs, $\theta$ is the fraction of the
total energy of the explosion converted into cosmic ray energy,
and $n$ and $d$ have their usual meaning, number density and
distance, respectively. In most cases, the expected flux at GeV
energies is far too low to be detected in the range of EGRET, but
the presence of nearby clouds can produce a significant
enhancement of the $\gamma$-ray emission. Such scenario has been
recently studied by Combi et al (1998) in relation with the source
3EG J1659-6251.

As showed by Aharonian et al. (1994), when a SNR hits a cloud a
part of the proton population can be transported into it by
convection, and illuminate the cloud by subsequent p-p
interaction, via pion decay. The resulting flux is \be  F{(\geq
100 {\rm MeV})} \sim \frac{q_\gamma M_{cloud}}{4 \pi d^2 m_H}, \ee
where $q_\gamma$ is the $\gamma$-ray emissivity per H-atom in the
cloud. This parameter can be related with its value in the
vicinity of Earth by $q_\gamma=k q_\odot$, with $k>1$.

Other mechanisms, like relativistic bremsstrahlung and inverse
Compton losses, associated with the leptonic component, can also
play a role if the electron density and/or the photon fields are
high enough, see for instance Pohl (1996). Then, there are many
theoretical reasons to expect a positional correlation between
SNRs and unidentified $\gamma$-ray sources. We devote the rest of
this work to analyze this correlation.

\section{Positional correlation}

Possible physical correlation between SNRs and unidentified EGRET
sources, on the basis of two dimensional positional coincidence,
has been proposed since the release of COS-B data (Montmerle 1979)
and the first EGRET (1EG) catalog. Sturner \& Dermer (1995)
suggested that some of the unidentified sources lying at low
galactic latitudes $|b|<10^o$ might be associated with SNRs: of 37
detections, 13 overlapped SNR positions in the 1EG catalog.
However, the statistical significance was not too high as to
provide a strong confidence. Using the 2EG catalog, Sturner et al.
(1996) repeated the analysis, and showed that 95\% confidence
contours of 7 unidentified EGRET sources overlapped SNRs, some of
them appearing to be in interaction with molecular clouds. Similar
results were independently reported by Esposito et al. (1996),
considering only radio-bright SNRs, and Yadigaroglu and Romani
(1997), although they did not assessed the overall chance
probability of these 2EG-catalog findings. The evolution in the
number of coincidences and SNRs considered in the different
studies is shown in Table~1. Note that from the First to the
Second EGRET catalogs, fourteen unidentified sources were
discarded. In the Third catalog, 6 unidentified sources are
possibly artifacts produced by the strong emission of the Vela
pulsar: these sources disappear in a map where the pulsed Vela
emission is suppressed.\footnote{Notes corresponding to Table 1
are as follow: a: Sturner and Dermer (1995). b: Sturner, Dermer
and Mattox (1996). c: Esposito, Hunter, Kanbach and Sreekumar
(1996). d: Yadigaroglu and Romani (1997). e: Romero, Benaglia and
Torres (1999). f: This work. g: Only radio-bright SNRs, flux at 1
GHz greater than 100 Jy, were used. h: Computed for pairs.}

In Table 2 we show the 3EG sources that are positionally
coincident with SNRs listed in the latest version of Green's
catalog. From left to right we provide the $\gamma$-ray source
name, the measured flux, the photon spectral index $\Gamma$, the
variability index $I$ (see below), the SNR identification, the
angular distance between the best $\gamma$-ray source position and
the center of the remnant, the size of the remnant in arcminutes,
the SNR type (S for shell, F for filled-centre, and C for
composite), and other positional coincidences found in our
previous study (Romero, Benaglia and Torres 1999). It is
interesting to note that in the 3EG catalog, not all the
positional coincidences with SNRs are SNOBs, as was the case
reported by Montmerle (1979) and Yadigaroglu and Romani (1997)
using previous samples.

\begin{figure}[ht]
\centerline{\epsfig{file=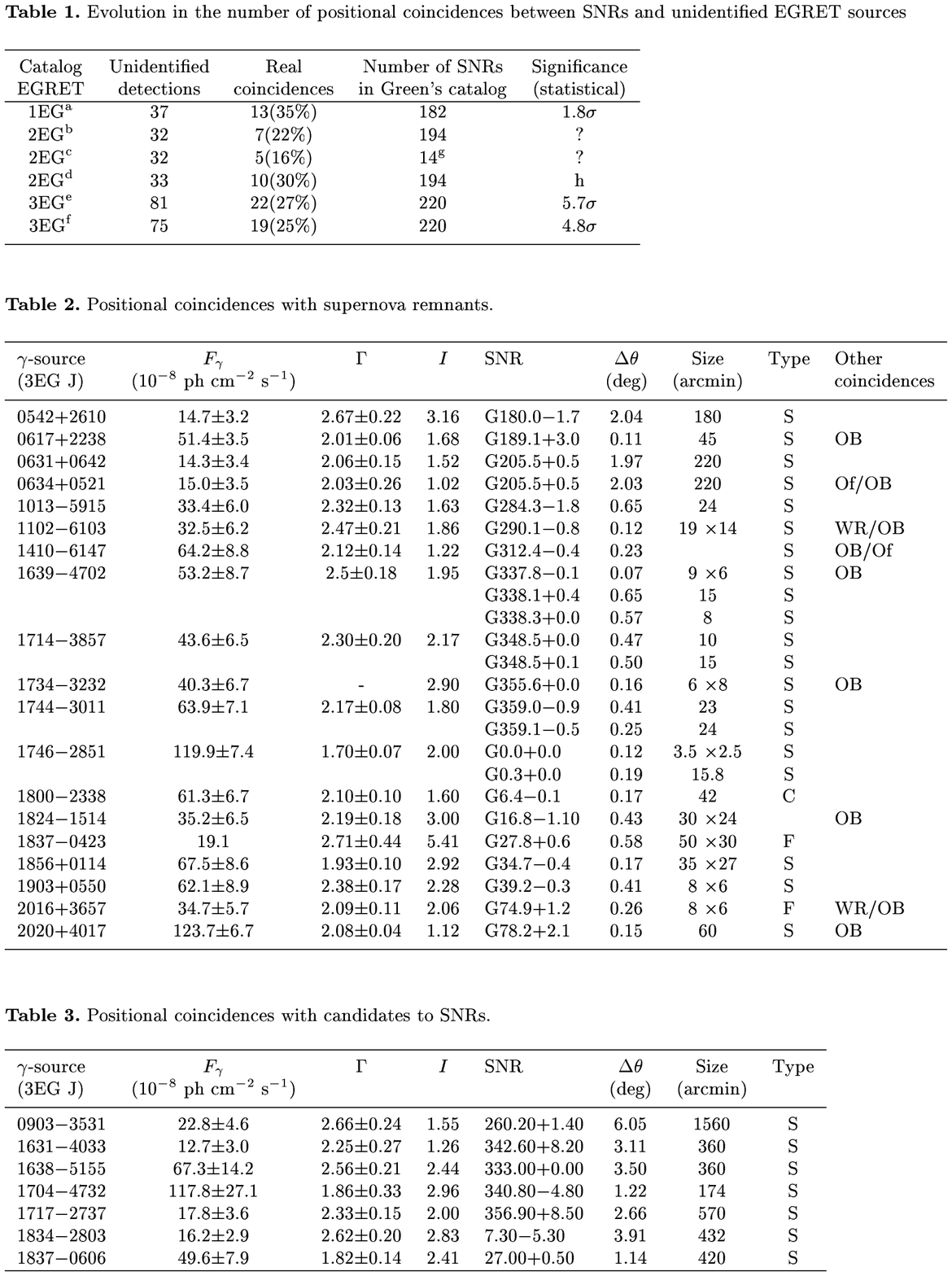, width=1.1\hsize}}
\end{figure}

We also consider whether some of the sources in our sample may be
associated with recently proposed candidates to supernova
remnants, presently not catalogued by Green (1998). Our interest
in this search resides in the fact that young stellar objects,
like recently formed black holes and pulsars, can still be
associated with the gaseous remnant of the original supernova that
created them. The diffuse non-thermal emission of the galactic
disk, originated in the interaction of the leptonic component of
the cosmic rays with the galactic magnetic field, is surely
veiling many remnants of low surface brightness. Recent
observational studies using filtering techniques in the analysis
of large-scale radio data have revealed several new SNR candidates
that are not yet included in the latest issue of Green's catalog.
In general, these new candidates are much more extended than those
previously known. There are 101 of these weak non-thermal
structures detected so far in the Galaxy. This number
significantly extends Green's (1998) catalogue. The list of these
new candidates, and the references from where they were compiled,
can be obtained from the paper by Torres et al. (2000).

We have found that only 7 gamma-ray sources in our sample are
positionally coincident with non-thermal radio structures. The
positional coincidences thus obtained are shown in Table~3, where
we provide similar information as that given in Table~2.

In order to estimate the statistical significance of these
coincidences, we have numerically simulated a large number of
synthetic sets of EGRET sources using the code described in the
paper by Romero, Benaglia and Torres (1999). The results of this
study are shown in Table~4 where we provide results for different
samples of unidentified sources and SNRs catalogs: the original 81
unidentified EGRET sources, the 75 high confidence ones, and the
40 sources without any positional correlation with known galactic
gamma-ray emitters (as reported by Romero et al. (2000)). For the
latter sample we are interested in checking the possible
association with candidates to SNRs. We conclude that there is
strong statistical evidence suggesting that some 3EG unidentified
EGRET sources must be physically associated with SNRs in Green's
Catalog. At the same time, there is no statistical evidence
suggesting that the 3EG sources analyzed in our sample are
physically associated with candidates to SNRs: the number of real
positional coincidences with them is totally compatible with, and
even lower than, the result of a random association. This latter
result appear to be the consequence of the candidates to SNRs
being much more extended objects, thus improving the random
coincidences with any given population, and does not discard in
itself that some of the coincidences could be physical ones.

\begin{table}
{\small {\bf Table  4.} Statistical results. G stands for Green's
SNRs and C for candidates.}
\begin{flushleft}
\begin{tabular}{l c c c c }
\noalign{\smallskip} \hline \noalign{\smallskip} Number &
Unidentified & Real & Simulated  & Poisson \cr of SNRs & Sources &
coincidences & coincidences&probability \cr \hline 220 (G) & 81 &
22 & 7.8 $\pm $ 2.5 & 1.5 $\times 10^{-5}$ \cr 220 (G) & 75 & 19 &
7.0 $\pm $ 2.4 & 9.9 $\times 10^{-5}$ \cr 321 (G $+$ C) & 75 & 30
& 22.4 $\pm $ 3.8 & 0.02 \cr 101 (C) & 40 & 7 & 10.4 $\pm $ 2.7 &
0.07 \cr \noalign{\smallskip} \hline
\end{tabular}
\end{flushleft}
\end{table}

\section{Spectral and variability indices}

In Figure 1 we present the distribution of the spectral index for
all unidentified sources which resulted to have positional
correlation with SNRs and candidates. All values are compatible
with Fermi-like acceleration processes that could happen in the
strong shocks generated by the explosion of the supernova, or be
the outcome of the interaction between relativistic material and a
nearby cloud, or --for the lowest ones-- be associated with
gamma-ray emission by pulsars. The distribution peaks around
$\Gamma = 2.1$, but some sources have a large spectral index of
about 2.8. Recall that the steepest measured spectral index for
pulsars is around 2.2, so this is pointing against a pulsar origin
for several of these gamma-ray sources.

\begin{figure}[ht]
\centerline{\epsfig{file=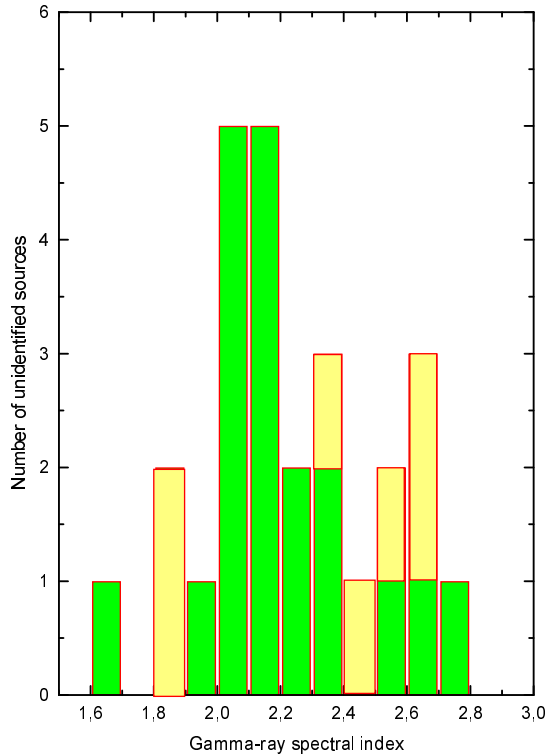, width=0.7\hsize}}
\caption{\protect\inx{Distribution of } spectral index for the
sources with positional coincidences with SNRs. The clearer boxes
stand for the sources coincident with candidates to SNRs.}
\label{fig.1}\end{figure}

We now assess the possible long term variability of the sources.
We define a mean weighted value for the EGRET flux as: \be \left<
F \right> = \left[ \sum_{i=1}^{N_{{\rm vp}}}
\frac{F(i)}{\epsilon(i)^2} \right]\times \left[
\sum_{i=1}^{N_{{\rm vp}}} \frac 1{\epsilon(i)^{2}}
\right]^{-1}.\ee $N_{{\rm vp}}$ is the number of viewing periods
for each gamma-ray source. $F(i)$ is the observed flux in the
$i^{{\rm th}}$-period, whereas $\epsilon(i)$ is the corresponding
error in the observed flux.\footnote{For those observations in
which the significance ($\sqrt{TS}$ in the EGRET catalog) is
greater than 3$\sigma$, we took the error as $\epsilon(i) =
F(i)/\sqrt{TS}$. However, many of the observations are in fact
upper bounds on the flux, with significance below 3$\sigma$. For
these ones, we assume both $F(i)$ and $\epsilon(i)$ as half the
value of the upper bound.} We then define the fluctuation index
$\mu$ as: $\mu =100\times \sigma_{{\rm sd}}\times \left< F
\right>^{-1}.$ In this expression, $\sigma_{{\rm sd}}$ is the
standard deviation of the flux measurements. In order to remove as
far as possible any spurious variability introduced by the
observing system, we computed the fluctuation index $\mu$ for the
confirmed gamma-ray pulsars in the 3EG catalog. The identification
by Kuiper et al. (2000) is not included because of the blazar
contamination of the EGRET flux, observed by these authors. We
adopt the physical criterion that pulsars are non-variable
gamma-ray sources. Then, any non-null $\mu$-value for pulsars is
attributed to experimental uncertainty. We then define an averaged
statistical index of variability, $I$, as \be I=\frac{\mu_{{\rm
source}}}{<\mu>_{{\rm pulsars}}}=\frac{\mu_{{\rm source}}}{26.9}.
\ee \mbox{} In terms of the averaged index $I$, the adopted
variability criterion is then that variable sources will be those
with $I>2.5$, which is 3$\sigma$ away from the statistical
variability of pulsars.\footnote{Detailed comments on how this
variability index compares with others, especially with the
analysis made in W. Tompkins' Ph.D. thesis, are given elsewhere
(Torres et al. 2000). }

In Figure 2 we show the distribution for the variability index $I$
of those sources that are positionally coincident with SNRs. We
see that most of the 19 sources positionally related with Green's
SNRs have $I<2.5$, being their mean value 2.17. Indeed, 12 sources
out of 19 have $I < 2$. There are three sources with very high
$I$-index: 3EG J1824-1514, J1837-0423 and 3EG J0542$+$2610, with
$I=3.00$, $I=5.41$ and $I=3.16$, respectively. These sources show
spectral indices of 2.19, 2.71 and 2.67. The source 3EG J1824-1514
has been recently proposed by Paredes et al. (2000) as a faint
microquasar detected through VLBI observations. This and other
sources with $I>2.5$, are very unlikely physically associated with
the SNRs in the usual sense, i.e. being pulsars or nearby clouds
in interaction with the swept up material. Nevertheless, compact
objects like accreting black holes, or even isolated Kerr-Newmann
black holes (Punsly et al. 2000) are interesting possibilities,
they surely requires additional analysis.

\begin{figure}[ht]
\centerline{\epsfig{file=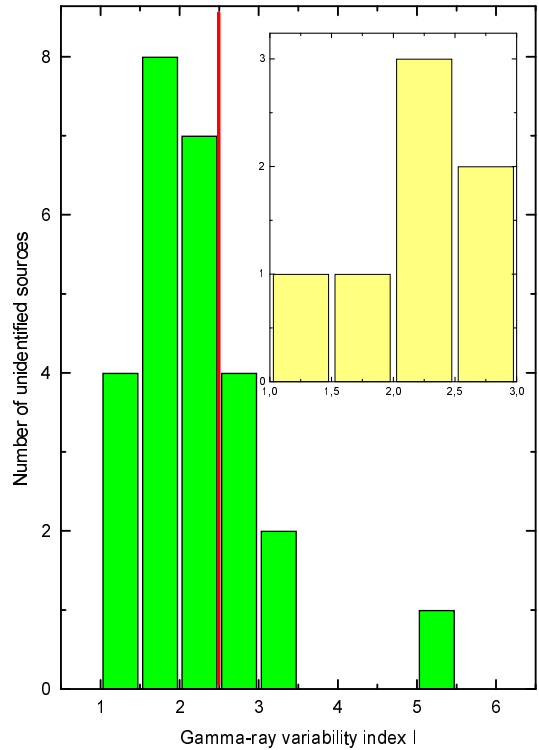, width=0.7\hsize}}
\caption{\protect\inx{Distribution of } the variability index for
the sources with positional coincidences with SNRs. The upper box
is the distribution only for the sources with positional
coincidences with candidates to SNRs. The vertical line separates
the sources which are variable (to the right) from the
rest.}\label{fig.2}\end{figure}

\begin{acknowledgments}
This work was partially supported by CONICET, ANPCT (PICT 98 No.
03-04881), and Fundaci\'{o}n Antorchas (through separate grants to
D.F.T., J.A.C. and G.E.R.). Additional on-line data can be
obtained at URL: www.iar.unlp.edu.ar/garra. Authors would like to
thank the Organizers for their kind invitation to this meeting.
\end{acknowledgments}

\end{document}